\documentclass{bmcart}

\input{glyphtounicode}
\usepackage[utf8]{inputenc} 

\usepackage{microtype}
    \DisableLigatures[f]{encoding=T1}
\usepackage[T1]{fontenc}

\usepackage{amsmath}
\usepackage{amssymb}
\usepackage{amsthm}

\usepackage{graphicx}
\usepackage{url}
\usepackage[hidelinks]{hyperref}
\usepackage{authblk}
\usepackage{booktabs}

\usepackage{enumitem}

\usepackage{changes}

\startlocaldefs

\newcommand{\DM} {\normalfont{\textsf{DM}}}
\newcommand{\dDM} {\normalfont{\textsf{dDM}}}
\newcommand{\dPDM} {\normalfont{\textsf{dPDM}}}

\newcommand{\calA}{\ensuremath{\mathcal{A}}}
\newcommand{\calB}{\ensuremath{\mathcal{B}}}
\newcommand{\calD}{\ensuremath{\mathcal{D}}}
\newcommand{\calL}{\ensuremath{\mathcal{L}}}
\newcommand{\calN}{\ensuremath{\mathcal{N}}}
\newcommand{\calR}{\ensuremath{\mathcal{R}}}
\newcommand{\calS}{\ensuremath{\mathcal{S}}}

\newcommand{\allB}{\ensuremath{\mathbf{B}}}
\newcommand{\allD}{\ensuremath{\mathbf{D}}}
\newcommand{\allS}{\ensuremath{\mathbf{S}}}
\newcommand{\allN}{\ensuremath{\mathbf{N}}}

\endlocaldefs

\theoremstyle{definition}
\newtheorem{definition}{Definition}[section]

\begin{document}

\begin{frontmatter}

\begin{fmbox}
\dochead{Research}

\title{Wibson Protocol for Secure Data Exchange and Batch Payments}

\author[
   addressref={aff1},                   
   email={daniel@wibson.org}   
]{\inits{DF}\fnm{Daniel} \snm{Fernandez}}
\author[
   addressref={aff2},
   email={futo@disarmista.com}
]{\inits{AF}\fnm{Ariel} \snm{Futoransky}}
\author[
   addressref={aff1},                   
   email={gustavo@wibson.org}   
]{\inits{GA}\fnm{Gustavo} \snm{Ajzenman}}
\author[
   addressref={aff1},                   
   email={mat@wibson.org}   
]{\inits{MT}\fnm{Matias} \snm{Travizano}}
\author[
   addressref={aff1},                   
   corref={aff1},                       
   email={charles@wibson.org}   
]{\inits{CS}\fnm{Carlos} \snm{Sarraute}}

\address[id=aff1]{
  \orgname{Wibson}, 
  \street{28 Irish Town},                     %
  \city{Gibraltar},                              
  \cny{UK}                                    
}
\address[id=aff2]{
  \orgname{Disarmista}, 
  \city{Buenos Aires},                              
  \cny{Argentina}                                    
}

\maketitle

\end{fmbox}

\begin{abstractbox}

\begin{abstract}

Wibson is a blockchain-based, decentralized data marketplace that provides
individuals a way to securely and anonymously sell information
in a trusted environment.
The combination of the Wibson token and blockchain-enabled smart contracts hopes to allow Data Sellers and Data Buyers to transact with each other directly while providing individuals the ability to maintain anonymity as desired.

The Wibson marketplace will provide infrastructure and financial incentives for individuals to securely sell personal information without sacrificing personal privacy. Data Buyers receive information from willing and actively participating individuals with the benefit of knowing that the personal information should be accurate and current.

We present here two different components working together to achieve an efficient decentralized marketplace. The first is a smart contract called Data Exchange, which stores references to Data Orders that different Buyers open in order to show to the market that they are interested in buying certain types of data, and provides secure mechanisms to perform the transactions. 
The second is used to process payments from Buyers to Sellers and intermediaries, and is called Batch Payments.

\end{abstract}


\end{abstractbox}

\end{frontmatter}


\section{Introduction}
\label{sec:introduction}

The Internet is a decentralized system that links billions of interconnected devices to improve communication, access to information, and economic possibilities for people across the globe. However, despite its distributed nature, giant technology companies (such as Google, Facebook, Apple, Microsoft and Amazon) have used the underlying technical protocols to build layers of proprietary applications that capture and control massive amounts of personal data.

Under the current system, individuals lack control over how data brokers collect, analyze, protect and use their personal data. Over the years, different revenue models have emerged evaluating how users can monetize their personal data~\cite{kemppainen2018emerging}. Additionally, governmental bodies and consumer rights organizations, especially in the European Union, are trying to maintain an appropriate balance between transparency, personal data use and data access.  
The prevailing data ecosystem misallocates data’s value away from the owner –the individual– and prevents society from effectively tackling many of its biggest challenges. Privacy advocates therefore claim that the time has come for citizens to regain control over their personal data and benefit from the value it creates, which sometimes results in new legislation~\cite{alix2018california}. 

In this paper, we are proposing a new market-based approach that leverages the latest developments in blockchain, cryptography and market design to connect data consumers (companies, organizations) and data owners (essentially, individuals).
Wibson is a blockchain-based, decentralized data marketplace that provides the infrastructure for individuals to securely and anonymously sell personal information that is validated for accuracy. Wibson is built on a set of core principles: {transparency, anonymity, fairness, censorship resistant, and the individual's control over the use of their personal information.}

The design and price of information in data markets is an active field of study~\cite{bergemann2018markets,bergemann2018design}. 
The perception of value of data has changed in the last years, while economists have  suggested that data should be considered as labor~\cite{arrieta2018should}.
In the Wibson marketplace described herein, citizens will be able to participate in an efficiently functioning, decentralized marketplace that provides both financial incentives and control over their personal information.

The essence of blockchain lies in its ability to support trusted transactions via networked computation in place of human control~\cite{zhao2016overview}.
By leveraging trust mechanisms arising from blockchain, the Wibson marketplace enables users to share their personal information without sacrificing privacy.

One of the biggest challenges that was presented after implementing the first version of the Wibson protocol~\cite{travizano2018wibson} was how to reduce the gas that was being paid due to the large number of transactions that were occurring between Buyers and Sellers. 
We decided to implement a separate smart contract that batches several payments and then performs a single transaction to pay multiple users simultaneously. This contract is called Batch Payments, and was introduced to reduce the gas costs associated with the operations in the Ethereum blockchain. In addition, part of the logic for paying Notaries was included in this new smart contract. As a consequence, the Data Exchange contract was also considerably modified -- its new version is presented in Sections~\ref{sec:data-exchange} and \ref{sec:protocol}.

The remainder of this paper is organized as follows. 
We recall the marketplace definitions in Section~\ref{sec:definitions}.
We define the participants of Wibson in Section~\ref{sec:participants} and
its main components in Section~\ref{sec:components}, including the message structure for all off-chain communications.
Section~\ref{sec:querying-system} explains the querying system for Buyers, and describes the Data Ontology and Data Order structure.
Section~\ref{sec:batch-payments} briefly introduces the Batch Payments system.
Section~\ref{sec:data-exchange} explains the mechanism used to perform atomic data exchange operations in the marketplace.
Section~\ref{sec:protocol} puts all the pieces together and describes the new version of the whole Wibson protocol, and Section~\ref{sec:conclusion} concludes the paper.


\section{Marketplace Definitions}
\label{sec:definitions}

We recall the notion of Decentralized, Privacy-Preserving Data Marketplace (\dPDM) which was introduced in \cite{travizano2018wibson}.

\begin{definition}
A \emph{Data Marketplace} (\DM) is a platform for the trade of information which provides:
\begin{itemize}
\item The infrastructure for a market whereby parties engage in exchange: sellers offering their data in exchange for money from buyers.
\item Allow any data tradeable item to be evaluated and valued.
\item Incentives for all players to be honest, and an enforcement system to take actions if a dishonest behavior is chosen.
\item Incentives for all players to ensure that data is trustworthy, and to provide quality data with the addition of an enforcement system to ensure this.
\end{itemize}
\end{definition}

\begin{definition}
A \emph{Decentralized Data Marketplace} (\dDM) is a {\DM} where:
\begin{itemize}
\item There is no central authority which regulates the participants of the market.
\item There is no central data repository. The users who generate the data are the owners of the data, and keep their data in their own devices having full control of their data assets. 
\item There is no central funds repository, therefore providing a trustless system where actors do not have to entrust their funds to a third party.
\end{itemize}
\end{definition}

\begin{definition}
A \emph{Decentralized Privacy-Preserving Data Marketplace} (\dPDM) is a {\dDM} which allows users to sell private information, while providing them the following privacy guarantees:
\begin{itemize}
\item Participants anonymity: the identity of the Sellers and Buyers is not revealed, without their consent. In particular, the identity of the Data Seller is not revealed to the Data Buyer, without the consent of the Data Seller.
\item Transparency over Data usage: the Data Seller has visibility on how his Data is used by the Buyer.
\item Control over Data usage: the Data Seller can modify the rights over its Data at any time.

\end{itemize}
\end{definition}


\section{Wibson Participants}
\label{sec:participants}

Wibson provides a marketplace inspired by the {\dPDM} definition, while allowing for trade-offs that are practical and necessary to ensure a market adapted to today's industry structure. 
Part of the protocol is executed on-chain, and part off-chain.

\subsection{Marketplace Participants}

The protocol is conducted by four types of participants: Data Seller (Seller), Data Buyer (Buyer), Notary and Delegate. 

\begin{figure}[ht]
	\centering
	{\includegraphics[width=0.9\linewidth]{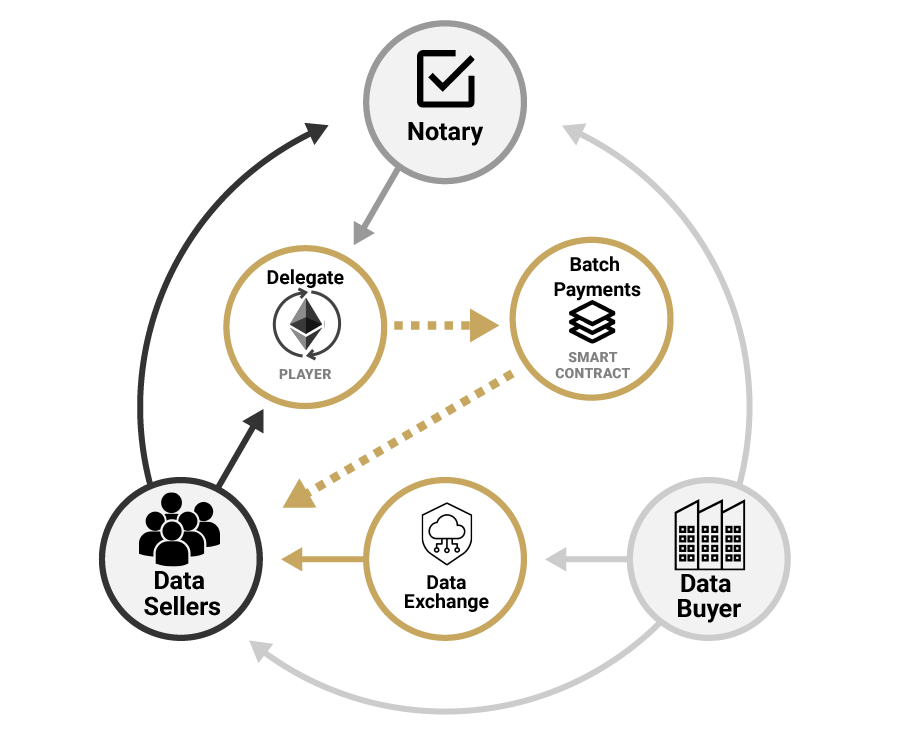}}
	\caption{The four type of Wibson participants are: Data Seller, Data Buyer, Notary and Delegate. The participants interact with two main components: the Data Exchange smart contract and the Batch Payments smart contract.}
	\label{fig:participants}
\end{figure}

\begin{description}
\item[Data Seller:] The Seller owns data and has rights to sell that data. The typical case is an individual selling his/her personal data. 

We denote the set of Sellers as 
$\allS = \{ \calS_1, \ldots, \calS_m \} $. 

\item[Data Buyer:] Any entity who wants to purchase data.

We denote the set of Buyers as 
$\allB = \{ \calB_1, \ldots, \calB_n \} $.

\item[Notary:] We introduce the role of Notary as a verification system to verify participants' information when required, verify data quality and trustworthiness when required, and arbitrate in case of conflict between 
Data Sellers and Data Buyers. 

We denote the set of Notaries as 
$\allN = \{ \calN_1, \ldots, \calN_p \} $. 

To qualify as a Notary, the Notary must have access to \emph{ground truth} information with respect to the data being exchanged in the marketplace. 
In other words, the Notary will be an entity with information from their own files on the Data Sellers and will be able to verify that information.

The Notary performs a verification on the Seller's data by comparing it with the ground truth data that the Notary owns. However, this does not imply that all the Seller's data should be transparent to the notary. For example, if the Notary is verifying the geolocation history of a Seller, the Notary may have only a subset of the geolocations of that Seller. The Notary can still verify that whenever there is a temporal match between his records and the ones provided by the Seller, the spatial distance between both records is reasonable (given the spatial resolution of the Notary's and the Seller's datasets).

We expect all Notaries to have public identities and an off-chain reputation.
We give in Section~\ref{sec:examples-participants} real-world examples of Notaries that can participate in the marketplace, and explain in each case the \emph{ground truth} information that enables them to act as Notaries.

\item[Delegate:] The Delegate is any entity that participates by sending transactions on behalf of another participants in exchange of earning an agreed fee. This new role was added in order to assist users that do not have Ether in their balances to submit transactions to Ethereum, and is part of the Batch Payments system.

We denote the set of Delegates as 
$\allD = \{ \calD_1, \ldots, \calD_q \} $. 

The role of the Delegate is to enable Sellers to participate in the marketplace without paying any gas. Thanks to the Delegate role, a Seller can enter the marketplace without paying any cost. The Delegate pays the gas costs of the Seller's transactions, and then receives a compensation in tokens for this service.
The role of the Delegate can be described as a gas abstraction mechanism. The Seller effectively pays a fee in tokens instead of gas for the transactions that he/she performs on the Ethereum blockchain.

\end{description}

\noindent
We say that the marketplace is \textbf{decentralized} since any participant which qualifies
can enter the marketplace as Data Seller, Data Buyer, Notary or Delegate.
There is no central authority which controls the participation in the market, or gives/denies permission to act in the market. 
 
There are other types of participants in the Batch Payments system, such as the Monitor who generates challenges, as a mechanism to ensure no fraudulent transactions are being made in the market (for more information see \cite{batpay2020}).

\subsection{Requirements for Participants of the Wibson Protocol}

\paragraph{Data Seller:}
In order to participate in the Wibson protocol, a Data Seller $\calS$ is required to have:
\begin{itemize}
\item Master Ethereum address to send and receive payments~\cite{buterin2014ethereum,wood2014ethereum}. 
The Seller may use different child addresses derived from their Ethereum master address for each transaction, thus hiding their other on-chain operations to other participants.
\item Public/private keys for signing transactions and encrypting data. 
\item Audience attributes (see Section~\ref{sec:querying-system}).
\item The Seller must be registered in the Batch Payments smart contract. Upon registration, he/she receives an identifier $\calS_{ID}$.
\end{itemize}

\paragraph{Data Buyer:}
In order to participate in the protocol, a Data Buyer $\calB$ is required to have:
\begin{itemize}
\item Ethereum address to send and receive payments.
\item Public/private keys for signing transactions and encrypting data.
\item Off-chain address to receive Data Responses. The off-chain address may be a URL or an IP address in the implementation of the protocol.
Alternatively, a decentralized peer-to-peer filesystem such as IPFS (InterPlanetary File System) can be used to share the data~\cite{ozyilmaz2018idmob}.
\item The Buyer must be registered in the Batch Payments smart contract. Upon registration, he/she receives an identifier $\calB_{ID}$.
\end{itemize}

\paragraph{Notary:} 
In order to participate in the protocol, a Notary $\calN$ is required to have:
\begin{itemize}
\item Ethereum address to send and receive payments.
\item Public/private keys for signing transactions and encrypting data.
\item Public URL address to receive data.
\item The Notary must be registered in the Batch Payments smart contract. Upon registration, he/she receives an identifier $\calN_{ID}$.
\item Mandatory: The Notary $\calN$ must reveal his public identity, by publishing his Ethereum address and public key in a publicly verifiable place.
\end{itemize}

The Notary must register himself on the Data Exchange smart contract, and publish his Public URL.
This establishes a link between the on-chain and off-chain worlds.

Additionally, in the Notary's public URL, he publishes his notarization fees (by type of data, per person) and terms of service.

\paragraph{Delegate:} 
In order to participate in the protocol, a Delegate $\calD$ is required to have:
\begin{itemize}
\item Ethereum address to send and receive payments.
\item Register in the Batch Payments smart contract. Upon registration, he/she receives an identifier $\calD_{ID}$.
\item The Delegate must have capital in order to operate in the Batch Payments system.
\end{itemize}

\subsection{Examples of Real World Market Participants}
\label{sec:examples-participants}

We illustrate the market roles with two examples.
 
\subsubsection{Bank Credit Card Transactions}

Suppose that the Seller is a client of a Bank, who offers on the market his (anonymized) credit card transactions. The Buyer can be any entity requiring transactional data to train its machine learning models. 

In this example, the Bank is the ideal Notary since:
\begin{itemize}
\item The Bank can verify that the Seller is actually a client of the Bank, by requiring the Seller to provide information that authenticates him/her.

This verification mechanism prevents a Sybil attack, wherein the attacker generates lot of Sellers. The Notary can detect whenever the same client is trying to sell multiple times the same data.

\item The Bank can act as a Notary in case of conflict, and verify whether the information of credit card transactions sent by the Seller to the Buyer is valid and trustworthy (in particular, by comparing with the Bank's own records of the client's credit card transactions).

\end{itemize}

\subsubsection{Location Data}

Suppose that the Seller is a client of a Telecommunications company, who offers on the market his (anonymized) records with location information. The Buyer can be any entity requiring location data to train its machine learning models. In this example, the Telecommunications company (Telco) is the ideal Notary since:
\begin{itemize}
\item The Telco can verify that the Seller is actually a client of the Telco, by requiring the Seller to provide information that authenticates him/her.

Here again, this verification mechanism prevents a Sybil attack, wherein the attacker generates lot of Sellers. The Notary can detect whenever the same client is trying to sell multiple times the same data.

\item The Telco can act as a Notary in case of conflict, and verify whether the location information sent by the Seller to the Buyer is valid and trustworthy (in particular, by comparing with the Telco's own records of the client's location when he used mobile phone services).

\end{itemize}


\section{Wibson Main Components}
\label{sec:components}
These are the main components of the Wibson protocol. We describe them in more detail in the following sections.

\begin{description}

\item[Querying System for Buyers.]
\textit{On-chain:} Buyers communicate their data requirements by placing Data Orders on the blockchain. We describe this system in Section~\ref{sec:querying-system}.

\item[Data Pricing Mechanism.]
\textit{On-chain:} The Buyer publishes on the blockchain the price offered for each Data Order. After selecting Data Responses, the Buyer publishes on the blockchain the selected Sellers, the price paid, and the hashes of the data.

\item[Payment System.] 
\textit{On-chain:} The initial implementation uses Ethereum smart-contracts~\cite{buterin2014ethereum,wood2014ethereum} and an ERC20 token~\cite{erc20}
using the Zeppelin’s implementation \emph{Standard Token}. The initial supply is 9,000,000,000 tokens with nine decimals.

The second part of the payment system is Batch Payments, which we describe in more detail in Section~\ref{sec:batch-payments}.

Ethereum is an open source, public, blockchain-based distributed computing platform and operating system featuring scripting functionality through smart contracts.
The Ethereum Virtual Machine (EVM) is the runtime environment for smart contracts in Ethereum.

Ethereum's smart contracts are based on different computer languages. Smart contracts are high-level programming abstractions that are compiled down to EVM bytecode and deployed to the Ethereum blockchain for execution. 
They can be written in Solidity (a language library with similarities to C and JavaScript).
Note that other blockchain platforms supporting smart contracts could be used to implement the marketplace protocol described in this paper. 

\item[Incentive System.]
The system provides mechanisms and incentives to certify participants, verifies that data is trustworthy, and incentivizes honest behavior of marketplace participants.

\textit{Off-chain:} Verification is performed by Notaries based on their proprietary information. Notaries audit transactions by signing them off-chain. The result of Notary audits is sent to the Buyer and used to determine payments to honest participants. Notaries earn tokens by verifying participants' information and validating data.

\textit{On-chain:} Unlock of payments is performed on-chain by the Notary. 

\item[Data Exchange.]
Wibson implements ideas presented in the Secure Exchange of Digital Goods~\cite{futoransky2019secure}. Section~\ref{sec:data-exchange} explains the mechanism to perform atomic data exchange operations.

\end{description}

\subsection*{Off-chain Messages}
\label{sec:offchain-messages}

Every message exchanged between parties in the Wibson Protocol must follow the Off-chain Message structure, comprised of:
\begin{enumerate}
\item The payload field: has the contents of the message. 
\item The signature field: contains the sender’s signature of the payload field.
\end{enumerate}

The message is encrypted with the Public Key of the recipient using Elliptic Curve Digital Signature Algorithm (ECDSA) before being sent~\cite{johnson2001elliptic}.


\section{Querying System for Buyers}
\label{sec:querying-system}

The Data Buyers communicate their data requirements by placing Data Orders on the blockchain. 
The Buyer indicates in the Data Order the intended audience and requested data, both specified using the Data Ontology. This section explains these concepts.

\subsection{Data Ontology for Audiences and Data Requests}
\label{sec:data-ontology}

\begin{definition}
The \textit{Data Ontology} is a publicly available document that formalizes naming, definition, structure and relationships~\cite{gruber1993translation} for the marketplace's data and can be used as a reference to generate Audiences and Data Requests. 

The Data Ontology is comprised of a comprehensive variable list that defines available \textit{Data Entities}, \textit{Data Query Models for each variable type}, and \textit{Audience Query Models} to filter available Data Sellers. Each particular implementation must define variables available in each category. 
\end{definition}

Given the publicly available definition of Data Ontology, a Data Buyer requests a particular Data Entity (e.g., browsing history) with additional parameters defined in the Data Query Model (e.g.,  two days of history) from an audience defined in the Audience Query Model (e.g., men who reside in Spain).

\begin{description}
\item[Data Entity:] Specific data owned by a Data Seller $\calS$. 
\begin{itemize}
\item Example 1: Browsing history. 
\item Example 2: Historical credit card transactions. 
\item Example 3: Seller's mobile phone Ad ID.
\end{itemize}

\item[Data Query Model:] A set of parameters that a Data Buyer $\calB$ can use to define the specific data amount, quality and type requested within a particular Data Entity. 
\begin{itemize}
\item Example 1: Two days of Data Seller browsing history beginning on January 1, 2017.
\item Example 2: All online credit card transactions.
\end{itemize}

\item[Audience Query Model:] A set of variables and values (or value ranges) that a Data Buyer $\calB$ can use to request data from relevant Data Sellers. 
\begin{itemize}
\item Example: Gender=Women, Age$\geq$40, Income$\geq$\$200,000, Current Residency=Spain.
\end{itemize}

\end{description}

\subsection{Data Order}

The Data Order (\textsf{DO}) is placed on-chain by the Data Buyer and includes: 
\begin{enumerate}[label=(\roman*)]
\item audience $\calA$,
\item requested data $\calR$, 
\item price $p$, 
\item hash of the terms and conditions of data use: $ H(\mbox{\textsf{tc}}) $,
\item public URL to upload Data Seller's responses and encrypted data via HTTPS post for this particular \textsf{DO}: $U_{\calB}$.

\end{enumerate}

\begin{description}
\item[Audience $\calA$:] The audience is a filter of potential sellers for the data order, written in terms of the Audience Query Model defined in a publicly published data ontology.
\begin{itemize}
\item Example: Gender=Women, Age$\geq$40, Income$\geq$\$200,000, Current Residency=Spain.
\end{itemize}

\item[Data Requested $\calR$:] The data requested is a list of Data Entities with certain parameters (defined in the Data Query Model), in addition to the Audience. 
The data requested can be empty. 
\begin{itemize}
\item Example 1: Credit card transactions over the last seven days. 
\item Example 2: Desktop browsing history over the last thirty days 
\item Example 3: Seller's mobile phone Ad ID.
\end{itemize}

\item[Address to Upload Data Seller's Responses and Data $U_{\calB}$:] Public URL address where the Buyer publishes additional information, and where the Buyer can receive data.
 
\end{description}

The public URL contains additional information about each specific Data Order:
\begin{enumerate}[label=(\roman*)]
\item the Data Buyer's public key $PK_{\calB}$,
\item the name of the Buyer,
\item a description of the Buyer,
\item the Buyer's logo,
\item the complete text of the terms and conditions of data use \mbox{\textsf{tc}}, whose hash $H(\mbox{\textsf{tc}})$ matches the hash published in the \textsf{DO},
\item intended use of data (chosen between the predefined categories),
\item list of accepted notaries for this \textsf{DO}, along with their fees, terms of services and signatures 
$\calL = \{ \calN_{1}, \ldots, \calN_{s} \}$. Buyers specify Notaries who are eligible to audit transactions, based on the match between Data Requested and Notary's verification capabilities.
\end{enumerate}


\section{Batch Payments}
\label{sec:batch-payments}

\emph{Batch Payments} is an intermediate smart contract to reduce gas costs associated with operating with existing ERC20 tokens on the Ethereum blockchain~\cite{batpay2020}.
It is a proxy scaling solution for the transfer of ERC20 tokens~\cite{erc20}. It is suitable for micropayments in one-to-many and few-to-many scenarios, including digital markets such as the Wibson market.

In BatPay (short for ``Batch Payments'') many similar operations are bundled together into a single
transaction in order to optimize gas consumption. In addition, some costly
verifications are replaced by a challenge game, pushing most of the computing cost
off-chain.
This results in a huge gas reduction in transfer costs.
In addition, it includes many relevant features, like meta-transactions for end-user
operation without ether, and key-locked transfers for atomic exchange of digital
goods.

We provide here a brief overview of the Batch Payments functionality.
These are the main characteristics:

\begin{enumerate}
\item Registration of all parties involved, where 32-bit account ids are used.
\item Buyers must add tokens to their BatPay account in order to make payments.
\item Buyers initiate payments by issuing a \textbf{registerPayment} transaction, which includes a
per-destination amount and a somewhat-compressed list of seller-ids. 
In this step, the accounts of the Buyers and Sellers are updated within the BatPay system. 
\item Sellers wait to accumulate enough payments, then initiate a \textbf{collect} transaction
specifying a range of payments and a total amount corresponding to their
account.
\item After a challenge period, the requested amount is added to the Seller's balance. The Seller may withdraw the tokens in his account after this step.
\item In the case of a dispute, the Seller lists the individual payments in which he/she is
included. The challenger selects a single payment and requests a proof of
inclusion. The loser pays for the verification game (stake).
\end{enumerate}

The version of the Wibson protocol using BatPay improves greatly the efficiency in which it registers or collects payments. We illustrate it with an example. Suppose that when a Data Buyer registers one payment for 1,000 Data Sellers, the transaction consumes 228,255 gas (229 gas each payment.) Furthermore, the gas consumption for the payment collection does not change with the number of payments. Collecting 1,000 payments consumes 167,440 gas (168 gas each payment), making a total of 397 gas per payment including both transactions, or \$0.00044 USD. Our solution improves efficiency by three orders of magnitude, since the transaction cost in the version without BatPay was \$0.45 USD for each piece of data exchanged on the market.


\section{Atomic Data Exchange}
\label{sec:data-exchange}

Here we propose a solution to the problem of trading real-world private information using cryptographic protocols and a public blockchain to guarantee the fairness of transactions.
This solution is inspired by the \emph{Secure Exchange of Digital Goods} (SEDG) protocol between a Data Buyer $\calB$ and a Data Seller $\calS$~\cite{futoransky2019secure}.
The protocol relies on a Notary $\calN$, which also plays the role of a trusted third party.

This protocol converts the exchange of data into an atomic transaction where three things happen simultaneously:
\begin{itemize}
\item The Buyer $\calB$ gets access to the Data, by learning the key that enables him to decrypt $C$ (previously received encrypted data).
\item The Seller $\calS$ gets paid for his Data when his key is revealed.
\item The Notary gets paid for his services at the very same time that the Seller gets paid for his data.
\end{itemize}

\begin{table}[ht]	
\caption{Atomic Exchange of Digital Goods}
\label{tab:sedg2}
{\normalsize 
\begin{center}
  \begin{tabular}[t]{ l l l }
    \toprule
Seller $\calS$  & Notary $\calN$	& Buyer $\calB$ \\ 
\midrule
	& $ M = \mbox{Random}() $ 		& \\ 
	& $ Lock = H( \calN_{ID} || M ) $ 		& \\
	& $ \mbox{Send}_{Buyer} ( Lock ) $	 & \\ \midrule
	$ K_{\calS} = \mbox{Random}() $ 		& & \\
	$ C_{\calS} = E_{K_{\calS}} (\mbox{Data}_{\calS}) $		& & \\
	$ \mbox{Send}_{Buyer} ( C_{\calS} ) $	& & \\ 
	$ \mbox{Send}_{Notary} ( C_{\calS}, K_{\calS} ) $	& & \\ 
\midrule
	& & $ h_{\calS} = H( C_{\calS} ) $		\\
	& & $ \mbox{Send}_{Notary} ( \calS_{ID}, h_{\calS} ) $	\\ 
\midrule
	& Verify $ H( C_{\calS} ) = h_\calS $ & \\
	& $ c_{K_{\calS}} = E_M( K_\calS ) $ 		& \\
	& $ \mbox{Send}_{Buyer} (c_{K_{\calS}}) $	 & \\ \midrule
	& & $ \mbox{RegisterPayment} ( \calS_{ID}, Lock ) $ \\ \midrule
	& $ \mbox{UnlockPayment} ( \calN_{ID}, M ) $	& \\
    \bottomrule
  \end{tabular}
\end{center}
}
\end{table}

This variation of the SEDG protocol allows the Notary to be paid for his services simultaneously with the Data Seller $\calS$. 
The solution is described in Table~\ref{tab:sedg2}.

The protocol begins with a \textbf{setup phase}, wherein the Notary generates a master key $M$ that will be used to encrypt all the Sellers' keys. He also generates a $ Lock = H( \calN_{ID} || M ) $ that is used by Batch Payments. To complete the setup for this transaction, he sends the $Lock$ to the Buyer.

In the following \textbf{transaction phase}, the Seller generates a random key $K_\calS$, that will be used to encrypt the data only for this transaction. Then the Seller sends to the Buyer his/her
data encrypted using the key $ K_\calS $: $ C_\calS = E_{K_\calS} (\mbox{Data}_{\calS}) $

The Seller also sends the encrypted data to the Notary, along with the encryption key $K_\calS$. The Notary can thus compute $ \mbox{Data}_{\calS} = D_{K_\calS} ( C_\calS ) $ and verify the data if required.

After the Buyer confirms that he is interested in the data of $\calS$, he sends the Notary the address of $\calS$ and $h_\calS$ (the hash of the Seller's encrypted data).

After the Notary performs the notarization and approves the data, he encrypts the Seller's key using the master key $M$. Here we present a simplified version, since the master key $M$ is used for the complete set of sellers approved in each transaction (the general setting is described in Section~\ref{sec:protocol}).
The Notary sends the Buyer
the key $ K_\calS $ encrypted with master key $ M $: $ c_{K_{\calS}} = E_M( K_\calS ) $.

The Data Buyer calls the \textbf{RegisterPayment} method of Batch Payments, specifying the address of the Seller, and the Lock which depends on the Notary's ID $\calN_{ID}$ and the master key $M$.

Finally, the Notary reveals the master key $ M $ by calling the UnlockPayment method.

When the Notary publishes the master key $ M $, the Batch Payments contract transfers the payments to the Seller and to the Notary. At the same time,
the Buyer can compute the following:
\begin{enumerate}[label=(\roman*)]
\item Verify the master key published $M'$ by checking that $ H ( \calN_{ID} || M' ) = Lock $.
\item Use $M'$ to compute $ K' = D_{M'} ( c_{K_{\calS}} ) $.
\item Use $K'$ to compute $ \mbox{Data} = D_{K'} ( C ) $.
\end{enumerate}
 
After the transaction is completed, the Buyer $\calB$ uses the master key to gain access to the Seller's data. This mechanism, wherein certain content is maintained private until a particular event (the publication of $M$) occurs, is reminiscent of the family of cryptographic primitives called Secure Triggers~\cite{futoransky2006foundations}.


\section{Wibson Protocol}
\label{sec:protocol}

\begin{figure}[ht]
	\centering
	{\includegraphics[width=\linewidth]{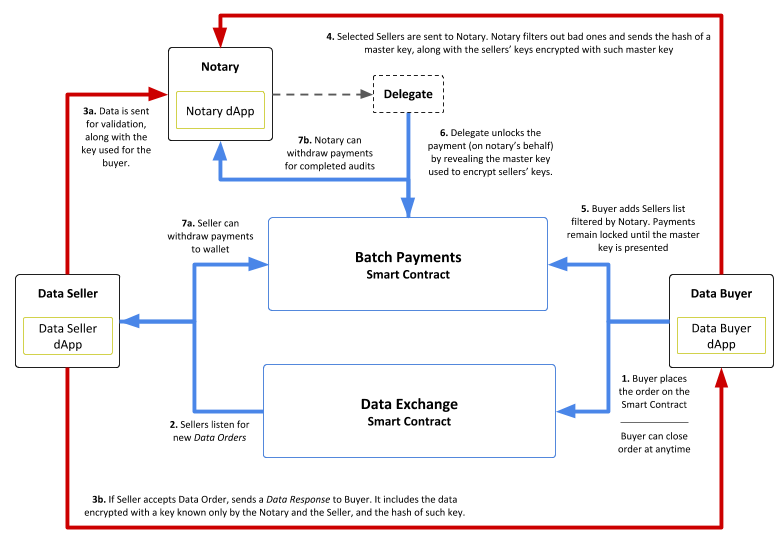}}
	\caption{Wibson's normal flow of operations.} 
	\label{fig:normal-flow}
\end{figure}

We present here the mechanisms and flow of operations of the Wibson protocol. 
The normal flow of operations is represented in Figure~\ref{fig:normal-flow}.

\begin{enumerate}

\item The Data Buyer $\calB$ creates a Data Order query 
$ \mbox{\textsf{DO}} = \left< \calA, \calR, p, H(\mbox{\textsf{tc}}), U_{\calB}  \right> $ 
in the Data Exchange, obtaining an ID for it in return.
The $ \mbox{\textsf{DO}}$ includes:
\begin{enumerate} 
\item audience $\calA$, 
\item requested data $\calR$, 
\item price $p$, 
\item hash of the terms and conditions of data use $ H(\mbox{\textsf{tc}}) $, 
\item public URL to upload Data Seller's responses and encrypted data via HTTPS post $U_{\calB}$ for this particular \textsf{DO}. 
The public URL contains additional information for this specific Data Order, including
the Data Buyer's public key $PK_{\calB}$
and the list $\calL = \{ \calN_{1}, \ldots, \calN_{s} \}$ of accepted notaries for this \textsf{DO}.

\end{enumerate}

The creation of the \textsf{DO} will emit an event so that everyone is aware of this new Data Order.

\medskip
\item Sellers monitor Data Orders and look for opportunities where they:
\begin{enumerate} 
\item match Data Order's audience $\calA$, 
\item agree on data requested $\calR$, 
\item accept the Data Order price $p$,
\item accept terms and conditions of data use \mbox{\textsf{tc}},
\item accept one of the suggested Notaries in $\calL$,
 \end{enumerate}

\medskip
\item If the Data Seller $\calS$ accepts a Data Order, he or she
sends a signed Data Response (Off-chain Message) directly to the Data Buyer’s public URL $U_\calB$, and an Off-chain Message directly to the Notary's public URL $U_\calN$.

To encrypt the data, the Seller generates a random key $ K_\calS $, that will be used only for this transaction.

The Data Response 
$\mbox{\textsf{DR}} = \left< \mbox{\textsf{DO}}_{ID}, \calS_{ID}, \calS_{eth}, \calN_{eth}, C_\calS, \mbox{\textsf{nbr}}  \right> $ that he sends to the Data Buyer includes:
\begin{enumerate} 
\item Data Order ID: $\mbox{\textsf{DO}}_{ID}$, 
\item Batch Payments Seller ID: $\calS_{ID}$, 
\item Ethereum Seller address $\calS_{eth}$ (if the Data Seller has no Batch Payments Seller ID), 
\item Notary Ethereum address $\calN_{eth}$. This is the selected Notary $\calN \in \calL$ who is included in the list of suggested Notaries $\calL$.
\item Data encrypted using the key $ K_\calS $: $ C_\calS = E_{K_\calS} (\mbox{Data}_{\calS}) $, 
\item NeedsBuyerRegistration \textsf{nbr}: boolean (for the Batch Payments registration). In the case that the Seller is not registered in Batch Payments, he can request the Buyer to register him automatically for this transaction.
\end{enumerate}

The Seller also sends the following Off-chain Message directly to the Notary's public URL:
\begin{enumerate} 

\item Data Order ID: $\mbox{\textsf{DO}}_{ID}$, 
\item Batch Payments Seller ID: $\calS_{ID}$, 
\item Ethereum Seller address $\calS_{eth}$ (if the Data Seller has no Batch Payments ID),
\item Data encrypted using the key $ K_\calS $: $ C_\calS = E_{K_\calS} (\mbox{Data}_{\calS}) $, 
\item Key $ K_\calS $ used to encrypt the data sent to the Data Buyer. 
\end{enumerate}

\medskip
\item The Data Buyer $\calB$ selects the set of Data Sellers $T$ from whom he wants to buy, and sends the list of Seller IDs or addresses to the Notary in the following form:
	\begin{enumerate}
	\item Data Order ID: $\mbox{\textsf{DO}}_{ID}$,
	\item Callback URL to receive the response from the Notary,
	\item A list of the sellers to be notarized (in the normal flow, these are all the sellers selected in $T$). 
	\end{enumerate}

	For each seller $\calS $ in the list $T$, it includes:

		\begin{enumerate}[label=(\roman*)]
		\item Batch Payments Seller ID: $ {\calS}_{ID} $,
		\item Seller Ethereum address $ {\calS}_{eth} $,
		\item Hash of the key $ K_{\calS} $: $ h_{\calS} = H(K_{\calS}) $.
		\end{enumerate}

\medskip
\item The Notary will then create a master key $M$ to encrypt and lock the data. 

The Notary builds a list of notarization results for the Seller's data. 
The list contains for each Seller $\calS$:
	\begin{enumerate}
	\item Seller ID: $\calS_{ID}$,
	\item Seller address: $\calS_{eth}$,
	\item Notarization result  specifying one of the following scenarios: \\
		(i) The data will not be notarized, \\
		(ii) The data was notarized and is valid (approved), \\
		(iii) The data was notarized and is invalid (rejected).

	\item Key $K_\calS$ encrypted with master key $M$: $ c_{K_\calS} = E_M( K_\calS ) $.
	\end{enumerate}
	
Then the Notary sends in response:
\begin{enumerate}
\item Data Order ID: $\mbox{\textsf{DO}}_{ID}$.
\item The list of Sellers containing the notarization results and encrypted keys.
	
\item Notarization fee for the service performed.
\item Notarization percentage applied (percentage of sellers to be notarized that were actually audited). The Notary decides which Data Sellers are notarized.

\item Notary Ethereum address: $\calN_{eth}$.
\item Hash of the pay data $\mbox{\textsf{payDataHash}}$ to be sent to Batch Payments (list of Seller IDs written in an  efficient way, see \cite{batpay2020})
\item Lock to be sent to Batch Payments of the form:
$ Lock = H ( \calN_{ID} || M ) $.
\end{enumerate}

\medskip
\item The Data Buyer $\calB$ calls the \textbf{registerPayment} method of Batch Payments to add the filtered list of Data Sellers (filtered by removing the rejected Sellers) using the following parameters:
\begin{enumerate}
\item Data Order ID: $\mbox{\textsf{DO}}_{ID}$, 
\item Batch Payments Seller IDs, 
\item $ Lock = H ( \calN_{ID} || M ) $,
\item Notarization fee for the service performed, 
\item Notary Ethereum address $\calN_{eth}$, 
\item Notary signature received in previous step. 
\end{enumerate}

\medskip
\item The Notary reveals the master key $M$ by calling the \textbf{unlockPayment} method on the Batch Payments contract using the following parameters:
\begin{enumerate}
\item Batch Id, 
\item Master Key $M$. 
\end{enumerate}

This triggers payments for the Sellers and the Notary.

\medskip
\item At any time, Data Sellers or Notaries can withdraw payments by calling the Batch Payments contract.

\end{enumerate}

\section{An Implementation of the Wibson Data Marketplace}
\label{sec:implementation-of-wibson}

\textbf{Decentralized Applications description.} 
We have implemented decentralized applications to operate on Wibson, one for each type of participant, which enable them to follow the entire protocol previously described.

\subsection{Data Seller dApp}

With this decentralized app, the Data Seller may perform any of the following \textbf{on-chain} operations:
\begin{itemize}
\item Listen to new \textit{Data Orders} placed on the blockchain.
\item Accept Data Orders with a new signed Data Response.
\item Check if a \textit{Data Response} was approved, i.e. selected to be bought.
\item Receive a payment for the sold data.
\item Keep track of the token balance in the wallet.
\end{itemize}

\noindent
As well as carry out the following \textbf{off-chain} operations:
\begin{itemize}
\item Store locally the data owned.
\item Integrate with external data sources such as social networks, other devices and so on.
\item Send the data to the buyer’s public URL when it corresponds.
\item Send \textit{Data Responses} to the buyer’s public URL when the user wants to sell.
\end{itemize}

\subsection{Data Buyer dApp}

With this decentralized app, the Data Buyer may perform any of the following \textbf{on-chain} operations:
\begin{itemize}
\item Place new Data Orders on the blockchain, specifying in each one of them the wanted audience.
\item Add to the contract the Data Responses that he or she is willing to buy.
\item Pay the Seller for the data.
\item Pay the fee to a Notary.
\item Keep track of the token balance in the wallet.
\item Sign transactions.
\end{itemize}

\noindent
As far as \textbf{off-chain} operations are concerned, the dApp also:
\begin{itemize}
\item Provides an API which the user may expose publicly in order to receive Data Responses or data itself.
\item Keeps track and shows the status for every Data Order placed and Data Responses received.
\end{itemize}

\subsection{Notary dApp}

With this decentralized app, the Notary may perform any of the following \textbf{on-chain} operations:
\begin{itemize}
\item Agree to act as Notary when requested by Data Buyer.
\item Sign transactions.
\item Keep track of the status of every transaction in which the notary is involved.
\item Receive the fee for every audit done.
\item Keep track of the token balance in the wallet.
\end{itemize}

\noindent
As well as carry out the following \textbf{off-chain} operations:
\begin{itemize}
\item Validate the information.
\item Sign audit results.
\end{itemize}


\section{Conclusion}
\label{sec:conclusion}

Our aim with the Wibson decentralized marketplace presented here is to restore the individuals' ownership over their personal information.
The Wibson protocol will benefit consumers by providing them the ability to control and monetize their personal information. It will also give access to high quality and verified data to organizations which need to train Machine Learning algorithms and models, as well as an explicit consumer consent mechanism which will be absolutely critical as new privacy regulations are coming into effect.

In addition to the marketplace protocol, Wibson also provides primitives to solve efficiently the problem of fair exchange~\cite{cleve1986limits} by providing an efficient zero-knowledge contingent payment mechanism~\cite{campanelli2017zero,sasson2014zerocash,ben2018scalable,bitansky2013succinct}, reminiscent of the \emph{secure triggers} cryptographic primitive~\cite{futoransky2006foundations}. 

By supporting the principles of transparency, anonymity, fairness and control, we believe that the Wibson marketplace will gain the people's trust 
needed to develop a vibrant data marketplace, that represents a fundamental change in the way organizations collect and use personal information for their data science and business analytics needs.


\begin{backmatter}

\section*{Declarations}

\subsection*{Availability of data and materials}

The datasets used and/or analyzed during the current study are available from the corresponding author on reasonable request.

\section*{Author's contributions}

   DF, AF and GA participated in the implementation of the protocol.
   All authors discussed the protocol design and contributed to the final manuscript.

\section*{Acknowledgements}

The authors thank Nicolás Ayala, Nahuel Santoalla, Agustín Modugno, Juan Carrillo and Martín Manelli for their work on the
implementation of the Wibson platform.


\bibliographystyle{bmc-mathphys} 

\bibliography{wibson}


\begin{thebibliography}{21}
\ifx \bisbn   \undefined \def \bisbn  #1{ISBN #1}\fi
\ifx \binits  \undefined \def \binits#1{#1}\fi
\ifx \bauthor  \undefined \def \bauthor#1{#1}\fi
\ifx \batitle  \undefined \def \batitle#1{#1}\fi
\ifx \bjtitle  \undefined \def \bjtitle#1{#1}\fi
\ifx \bvolume  \undefined \def \bvolume#1{\textbf{#1}}\fi
\ifx \byear  \undefined \def \byear#1{#1}\fi
\ifx \bissue  \undefined \def \bissue#1{#1}\fi
\ifx \bfpage  \undefined \def \bfpage#1{#1}\fi
\ifx \blpage  \undefined \def \blpage #1{#1}\fi
\ifx \burl  \undefined \def \burl#1{\textsf{#1}}\fi
\ifx \doiurl  \undefined \def \doiurl#1{\textsf{#1}}\fi
\ifx \betal  \undefined \def \betal{\textit{et al.}}\fi
\ifx \binstitute  \undefined \def \binstitute#1{#1}\fi
\ifx \binstitutionaled  \undefined \def \binstitutionaled#1{#1}\fi
\ifx \bctitle  \undefined \def \bctitle#1{#1}\fi
\ifx \beditor  \undefined \def \beditor#1{#1}\fi
\ifx \bpublisher  \undefined \def \bpublisher#1{#1}\fi
\ifx \bbtitle  \undefined \def \bbtitle#1{#1}\fi
\ifx \bedition  \undefined \def \bedition#1{#1}\fi
\ifx \bseriesno  \undefined \def \bseriesno#1{#1}\fi
\ifx \blocation  \undefined \def \blocation#1{#1}\fi
\ifx \bsertitle  \undefined \def \bsertitle#1{#1}\fi
\ifx \bsnm \undefined \def \bsnm#1{#1}\fi
\ifx \bsuffix \undefined \def \bsuffix#1{#1}\fi
\ifx \bparticle \undefined \def \bparticle#1{#1}\fi
\ifx \barticle \undefined \def \barticle#1{#1}\fi
\ifx \bconfdate \undefined \def \bconfdate #1{#1}\fi
\ifx \botherref \undefined \def \botherref #1{#1}\fi
\ifx \url \undefined \def \url#1{\textsf{#1}}\fi
\ifx \bchapter \undefined \def \bchapter#1{#1}\fi
\ifx \bbook \undefined \def \bbook#1{#1}\fi
\ifx \bcomment \undefined \def \bcomment#1{#1}\fi
\ifx \oauthor \undefined \def \oauthor#1{#1}\fi
\ifx \citeauthoryear \undefined \def \citeauthoryear#1{#1}\fi
\ifx \endbibitem  \undefined \def \endbibitem {}\fi
\ifx \bconflocation  \undefined \def \bconflocation#1{#1}\fi
\ifx \arxivurl  \undefined \def \arxivurl#1{\textsf{#1}}\fi
\csname PreBibitemsHook\endcsname

\bibitem{kemppainen2018emerging}
\begin{barticle}
\bauthor{\bsnm{Kemppainen}, \binits{L.}},
\bauthor{\bsnm{Koivum{\"a}ki}, \binits{T.}},
\bauthor{\bsnm{Pikkarainen}, \binits{M.}},
\bauthor{\bsnm{Poikola}, \binits{A.}}:
\batitle{Emerging revenue models for personal data platform operators: When
  individuals are in control of their data}.
\bjtitle{Journal of Business Models}
\bvolume{6}(\bissue{3}),
\bfpage{79}--\blpage{105}
(\byear{2018})
\end{barticle}
\endbibitem

\bibitem{alix2018california}
\begin{barticle}
\bauthor{\bsnm{Alix}, \binits{L.}}:
\batitle{California passes nation’s first statewide consumer privacy law}.
\bjtitle{American Banker}
\bvolume{183}(\bissue{126}),
\bfpage{1}
(\byear{2018})
\end{barticle}
\endbibitem

\bibitem{bergemann2018markets}
\begin{botherref}
\oauthor{\bsnm{Bergemann}, \binits{D.}},
\oauthor{\bsnm{Bonatti}, \binits{A.}}, et al.:
Markets for information: An introduction.
Technical report,
Cowles Foundation for Research in Economics, Yale University
(2018)
\end{botherref}
\endbibitem

\bibitem{bergemann2018design}
\begin{barticle}
\bauthor{\bsnm{Bergemann}, \binits{D.}},
\bauthor{\bsnm{Bonatti}, \binits{A.}},
\bauthor{\bsnm{Smolin}, \binits{A.}}:
\batitle{The design and price of information}.
\bjtitle{American Economic Review}
\bvolume{108}(\bissue{1}),
\bfpage{1}--\blpage{48}
(\byear{2018})
\end{barticle}
\endbibitem

\bibitem{arrieta2018should}
\begin{bchapter}
\bauthor{\bsnm{Arrieta-Ibarra}, \binits{I.}},
\bauthor{\bsnm{Goff}, \binits{L.}},
\bauthor{\bsnm{Jim{\'e}nez-Hern{\'a}ndez}, \binits{D.}},
\bauthor{\bsnm{Lanier}, \binits{J.}},
\bauthor{\bsnm{Weyl}, \binits{E.G.}}:
\bctitle{Should we treat data as labor? moving beyond "free"}.
In: \bbtitle{AEA Papers and Proceedings},
vol. \bseriesno{108},
pp. \bfpage{38}--\blpage{42}
(\byear{2018})
\end{bchapter}
\endbibitem

\bibitem{zhao2016overview}
\begin{botherref}
\oauthor{\bsnm{Zhao}, \binits{J.L.}},
\oauthor{\bsnm{Fan}, \binits{S.}},
\oauthor{\bsnm{Yan}, \binits{J.}}:
Overview of business innovations and research opportunities in blockchain and
  introduction to the special issue.
Financial Innovation
\textbf{2}(1)
(2016).
doi:\doiurl{10.1186/s40854-016-0049-2}
\end{botherref}
\endbibitem

\bibitem{travizano2018wibson}
\begin{bchapter}
\bauthor{\bsnm{Travizano}, \binits{M.}},
\bauthor{\bsnm{Sarraute}, \binits{C.}},
\bauthor{\bsnm{Ajzenman}, \binits{G.}},
\bauthor{\bsnm{Minnoni}, \binits{M.}}:
\bctitle{Wibson: A decentralized data marketplace}.
In: \bbtitle{Proceedings of SIGBPS 2018 Workshop on Blockchain and Smart
  Contract}
(\byear{2018})
\end{bchapter}
\endbibitem

\bibitem{batpay2020}
\begin{botherref}
\oauthor{\bsnm{Mayer}, \binits{H.}},
\oauthor{\bsnm{Bejarano}, \binits{I.}},
\oauthor{\bsnm{Fernandez}, \binits{D.}},
\oauthor{\bsnm{Ajzenman}, \binits{G.}},
\oauthor{\bsnm{Ayala}, \binits{N.}},
\oauthor{\bsnm{Santoalla}, \binits{N.}},
\oauthor{\bsnm{Sarraute}, \binits{C.}},
\oauthor{\bsnm{Futoransky}, \binits{A.}}:
{B}at{P}ay: a gas efficient protocol for the recurrent micropayment of {ERC20}
  tokens
(2020)
\end{botherref}
\endbibitem

\bibitem{buterin2014ethereum}
\begin{botherref}
\oauthor{\bsnm{Buterin}, \binits{V.}}:
Ethereum: A next-generation smart contract and decentralized application
  platform.
Ethereum White Paper
(2014)
\end{botherref}
\endbibitem

\bibitem{wood2014ethereum}
\begin{botherref}
\oauthor{\bsnm{Wood}, \binits{G.}}:
Ethereum: A secure decentralised generalised transaction ledger.
Ethereum Project Yellow Paper
\textbf{151}
(2014)
\end{botherref}
\endbibitem

\bibitem{ozyilmaz2018idmob}
\begin{bchapter}
\bauthor{\bsnm{{\"O}zyilmaz}, \binits{K.R.}},
\bauthor{\bsnm{Do{\u{g}}an}, \binits{M.}},
\bauthor{\bsnm{Yurdakul}, \binits{A.}}:
\bctitle{Idmob: Iot data marketplace on blockchain}.
In: \bbtitle{2018 Crypto Valley Conference on Blockchain Technology (CVCBT)},
pp. \bfpage{11}--\blpage{19}
(\byear{2018}).
\bcomment{IEEE}
\end{bchapter}
\endbibitem

\bibitem{erc20}
\begin{botherref}
\oauthor{\bsnm{Vogelsteller}, \binits{F.}},
\oauthor{\bsnm{Buterin}, \binits{V.}}:
ERC-20 Token Standard.
\url{https://github.com/ethereum/EIPs/blob/master/EIPS/eip-20-token-standard.md}.
Accessed: 2018-01-25
\end{botherref}
\endbibitem

\bibitem{futoransky2019secure}
\begin{bchapter}
\bauthor{\bsnm{Futoransky}, \binits{A.}},
\bauthor{\bsnm{Sarraute}, \binits{C.}},
\bauthor{\bsnm{Waissbein}, \binits{A.}},
\bauthor{\bsnm{Travizano}, \binits{M.}},
\bauthor{\bsnm{Minnoni}, \binits{M.}}:
\bctitle{Secure exchange of digital goods in a decentralized data marketplace}.
In: \bbtitle{Proceedings of the 2019 Argentine Symposium on Big Data
  (AGRANDA)},
pp. \bfpage{38}--\blpage{44}
(\byear{2019})
\end{bchapter}
\endbibitem

\bibitem{johnson2001elliptic}
\begin{barticle}
\bauthor{\bsnm{Johnson}, \binits{D.}},
\bauthor{\bsnm{Menezes}, \binits{A.}},
\bauthor{\bsnm{Vanstone}, \binits{S.}}:
\batitle{The elliptic curve digital signature algorithm (ecdsa)}.
\bjtitle{International journal of information security}
\bvolume{1}(\bissue{1}),
\bfpage{36}--\blpage{63}
(\byear{2001})
\end{barticle}
\endbibitem

\bibitem{gruber1993translation}
\begin{barticle}
\bauthor{\bsnm{Gruber}, \binits{T.R.}}:
\batitle{A translation approach to portable ontology specifications}.
\bjtitle{Knowledge acquisition}
\bvolume{5}(\bissue{2}),
\bfpage{199}--\blpage{220}
(\byear{1993})
\end{barticle}
\endbibitem

\bibitem{futoransky2006foundations}
\begin{barticle}
\bauthor{\bsnm{Futoransky}, \binits{A.}},
\bauthor{\bsnm{Kargieman}, \binits{E.}},
\bauthor{\bsnm{Sarraute}, \binits{C.}},
\bauthor{\bsnm{Waissbein}, \binits{A.}}:
\batitle{Foundations and applications for secure triggers}.
\bjtitle{ACM Transactions on Information and System Security (TISSEC)}
\bvolume{9}(\bissue{1}),
\bfpage{94}--\blpage{112}
(\byear{2006})
\end{barticle}
\endbibitem

\bibitem{cleve1986limits}
\begin{bchapter}
\bauthor{\bsnm{Cleve}, \binits{R.}}:
\bctitle{Limits on the security of coin flips when half the processors are
  faulty}.
In: \bbtitle{Proceedings of the Eighteenth Annual ACM Symposium on Theory of
  Computing},
pp. \bfpage{364}--\blpage{369}
(\byear{1986}).
\bcomment{ACM}
\end{bchapter}
\endbibitem

\bibitem{campanelli2017zero}
\begin{bchapter}
\bauthor{\bsnm{Campanelli}, \binits{M.}},
\bauthor{\bsnm{Gennaro}, \binits{R.}},
\bauthor{\bsnm{Goldfeder}, \binits{S.}},
\bauthor{\bsnm{Nizzardo}, \binits{L.}}:
\bctitle{Zero-knowledge contingent payments revisited: Attacks and payments for
  services}.
In: \bbtitle{Proceedings of the 2017 ACM SIGSAC Conference on Computer and
  Communications Security},
pp. \bfpage{229}--\blpage{243}
(\byear{2017}).
\bcomment{ACM}
\end{bchapter}
\endbibitem

\bibitem{sasson2014zerocash}
\begin{bchapter}
\bauthor{\bsnm{Ben-Sasson}, \binits{E.}},
\bauthor{\bsnm{Chiesa}, \binits{A.}},
\bauthor{\bsnm{Garman}, \binits{C.}},
\bauthor{\bsnm{Green}, \binits{M.}},
\bauthor{\bsnm{Miers}, \binits{I.}},
\bauthor{\bsnm{Tromer}, \binits{E.}},
\bauthor{\bsnm{Virza}, \binits{M.}}:
\bctitle{Zerocash: Decentralized anonymous payments from bitcoin}.
In: \bbtitle{2014 IEEE Symposium on Security and Privacy (SP)},
pp. \bfpage{459}--\blpage{474}
(\byear{2014}).
\bcomment{IEEE}
\end{bchapter}
\endbibitem

\bibitem{ben2018scalable}
\begin{barticle}
\bauthor{\bsnm{Ben-Sasson}, \binits{E.}},
\bauthor{\bsnm{Bentov}, \binits{I.}},
\bauthor{\bsnm{Horesh}, \binits{Y.}},
\bauthor{\bsnm{Riabzev}, \binits{M.}}:
\batitle{Scalable, transparent, and post-quantum secure computational
  integrity}.
\bjtitle{Cryptol. ePrint Arch., Tech. Rep}
\bvolume{46},
\bfpage{2018}
(\byear{2018})
\end{barticle}
\endbibitem

\bibitem{bitansky2013succinct}
\begin{bchapter}
\bauthor{\bsnm{Bitansky}, \binits{N.}},
\bauthor{\bsnm{Chiesa}, \binits{A.}},
\bauthor{\bsnm{Ishai}, \binits{Y.}},
\bauthor{\bsnm{Paneth}, \binits{O.}},
\bauthor{\bsnm{Ostrovsky}, \binits{R.}}:
\bctitle{Succinct non-interactive arguments via linear interactive proofs}.
In: \bbtitle{Theory of Cryptography},
pp. \bfpage{315}--\blpage{333}.
\bpublisher{Springer},
\blocation{Berlin, Heidelberg}
(\byear{2013})
\end{bchapter}
\endbibitem

\end{thebibliography}

\newcommand{\BMCxmlcomment}[1]{}

\BMCxmlcomment{

<refgrp>

<bibl id="B1">
  <title><p>Emerging Revenue Models for Personal Data Platform Operators: When
  Individuals are in Control of Their Data</p></title>
  <aug>
    <au><snm>Kemppainen</snm><fnm>L</fnm></au>
    <au><snm>Koivum{\"a}ki</snm><fnm>T</fnm></au>
    <au><snm>Pikkarainen</snm><fnm>M</fnm></au>
    <au><snm>Poikola</snm><fnm>A</fnm></au>
  </aug>
  <source>Journal of Business Models</source>
  <pubdate>2018</pubdate>
  <volume>6</volume>
  <issue>3</issue>
  <fpage>79</fpage>
  <lpage>-105</lpage>
</bibl>

<bibl id="B2">
  <title><p>California Passes Nation’s First Statewide Consumer Privacy
  Law</p></title>
  <aug>
    <au><snm>Alix</snm><fnm>L</fnm></au>
  </aug>
  <source>American Banker</source>
  <pubdate>2018</pubdate>
  <volume>183</volume>
  <issue>126</issue>
  <fpage>1</fpage>
</bibl>

<bibl id="B3">
  <title><p>Markets for Information: An Introduction</p></title>
  <aug>
    <au><snm>Bergemann</snm><fnm>D</fnm></au>
    <au><snm>Bonatti</snm><fnm>A</fnm></au>
    <au><cnm>others</cnm></au>
  </aug>
  <pubdate>2018</pubdate>
</bibl>

<bibl id="B4">
  <title><p>The design and price of information</p></title>
  <aug>
    <au><snm>Bergemann</snm><fnm>D</fnm></au>
    <au><snm>Bonatti</snm><fnm>A</fnm></au>
    <au><snm>Smolin</snm><fnm>A</fnm></au>
  </aug>
  <source>American Economic Review</source>
  <pubdate>2018</pubdate>
  <volume>108</volume>
  <issue>1</issue>
  <fpage>1</fpage>
  <lpage>-48</lpage>
</bibl>

<bibl id="B5">
  <title><p>Should We Treat Data as Labor? Moving beyond "Free"</p></title>
  <aug>
    <au><snm>Arrieta Ibarra</snm><fnm>I</fnm></au>
    <au><snm>Goff</snm><fnm>L</fnm></au>
    <au><snm>Jim{\'e}nez Hern{\'a}ndez</snm><fnm>D</fnm></au>
    <au><snm>Lanier</snm><fnm>J</fnm></au>
    <au><snm>Weyl</snm><fnm>EG</fnm></au>
  </aug>
  <source>AEA Papers and Proceedings</source>
  <pubdate>2018</pubdate>
  <volume>108</volume>
  <fpage>38</fpage>
  <lpage>-42</lpage>
</bibl>

<bibl id="B6">
  <title><p>Overview of business innovations and research opportunities in
  blockchain and introduction to the special issue</p></title>
  <aug>
    <au><snm>Zhao</snm><fnm>JL</fnm></au>
    <au><snm>Fan</snm><fnm>S</fnm></au>
    <au><snm>Yan</snm><fnm>J</fnm></au>
  </aug>
  <source>Financial Innovation</source>
  <publisher>Springer Nature</publisher>
  <pubdate>2016</pubdate>
  <volume>2</volume>
  <issue>1</issue>
</bibl>

<bibl id="B7">
  <title><p>Wibson: A Decentralized Data Marketplace</p></title>
  <aug>
    <au><snm>Travizano</snm><fnm>M</fnm></au>
    <au><snm>Sarraute</snm><fnm>C</fnm></au>
    <au><snm>Ajzenman</snm><fnm>G</fnm></au>
    <au><snm>Minnoni</snm><fnm>M</fnm></au>
  </aug>
  <source>Proceedings of SIGBPS 2018 Workshop on Blockchain and Smart
  Contract</source>
  <pubdate>2018</pubdate>
</bibl>

<bibl id="B8">
  <title><p>{B}at{P}ay: a gas efficient protocol for the recurrent micropayment
  of {ERC20} tokens</p></title>
  <aug>
    <au><snm>Mayer</snm><fnm>H</fnm></au>
    <au><snm>Bejarano</snm><fnm>I</fnm></au>
    <au><snm>Fernandez</snm><fnm>D</fnm></au>
    <au><snm>Ajzenman</snm><fnm>G</fnm></au>
    <au><snm>Ayala</snm><fnm>N</fnm></au>
    <au><snm>Santoalla</snm><fnm>N</fnm></au>
    <au><snm>Sarraute</snm><fnm>C</fnm></au>
    <au><snm>Futoransky</snm><fnm>A</fnm></au>
  </aug>
  <pubdate>2020</pubdate>
</bibl>

<bibl id="B9">
  <title><p>Ethereum: A next-generation smart contract and decentralized
  application platform</p></title>
  <aug>
    <au><snm>Buterin</snm><fnm>V</fnm></au>
  </aug>
  <source>\url{https://github.com/ethereum/wiki/wiki/White-Paper}</source>
  <pubdate>2014</pubdate>
</bibl>

<bibl id="B10">
  <title><p>Ethereum: A secure decentralised generalised transaction
  ledger</p></title>
  <aug>
    <au><snm>Wood</snm><fnm>G</fnm></au>
  </aug>
  <source>Ethereum Project Yellow Paper</source>
  <pubdate>2014</pubdate>
  <volume>151</volume>
</bibl>

<bibl id="B11">
  <title><p>IDMoB: IoT Data Marketplace on Blockchain</p></title>
  <aug>
    <au><snm>{\"O}zyilmaz</snm><fnm>KR</fnm></au>
    <au><snm>Do{\u{g}}an</snm><fnm>M</fnm></au>
    <au><snm>Yurdakul</snm><fnm>A</fnm></au>
  </aug>
  <source>2018 Crypto Valley Conference on Blockchain Technology
  (CVCBT)</source>
  <pubdate>2018</pubdate>
  <fpage>11</fpage>
  <lpage>-19</lpage>
</bibl>

<bibl id="B12">
  <title><p>ERC-20 Token Standard</p></title>
  <aug>
    <au><snm>Vogelsteller</snm><fnm>F</fnm></au>
    <au><snm>Buterin</snm><fnm>V</fnm></au>
  </aug>
  <source>\url{https://github.com/ethereum/EIPs/blob/master/EIPS/eip-20-token-standard.md}</source>
  <note>Accessed: 2018-01-25</note>
</bibl>

<bibl id="B13">
  <title><p>Secure Exchange of Digital Goods in a Decentralized Data
  Marketplace</p></title>
  <aug>
    <au><snm>Futoransky</snm><fnm>A</fnm></au>
    <au><snm>Sarraute</snm><fnm>C</fnm></au>
    <au><snm>Waissbein</snm><fnm>A</fnm></au>
    <au><snm>Travizano</snm><fnm>M</fnm></au>
    <au><snm>Minnoni</snm><fnm>M</fnm></au>
  </aug>
  <source>Proceedings of the 2019 Argentine Symposium on Big Data
  (AGRANDA)</source>
  <pubdate>2019</pubdate>
  <fpage>38</fpage>
  <lpage>-44</lpage>
</bibl>

<bibl id="B14">
  <title><p>The elliptic curve digital signature algorithm (ECDSA)</p></title>
  <aug>
    <au><snm>Johnson</snm><fnm>D</fnm></au>
    <au><snm>Menezes</snm><fnm>A</fnm></au>
    <au><snm>Vanstone</snm><fnm>S</fnm></au>
  </aug>
  <source>International journal of information security</source>
  <publisher>Springer</publisher>
  <pubdate>2001</pubdate>
  <volume>1</volume>
  <issue>1</issue>
  <fpage>36</fpage>
  <lpage>-63</lpage>
</bibl>

<bibl id="B15">
  <title><p>A translation approach to portable ontology
  specifications</p></title>
  <aug>
    <au><snm>Gruber</snm><fnm>TR</fnm></au>
  </aug>
  <source>Knowledge acquisition</source>
  <publisher>Elsevier</publisher>
  <pubdate>1993</pubdate>
  <volume>5</volume>
  <issue>2</issue>
  <fpage>199</fpage>
  <lpage>-220</lpage>
</bibl>

<bibl id="B16">
  <title><p>Foundations and applications for secure triggers</p></title>
  <aug>
    <au><snm>Futoransky</snm><fnm>A</fnm></au>
    <au><snm>Kargieman</snm><fnm>E</fnm></au>
    <au><snm>Sarraute</snm><fnm>C</fnm></au>
    <au><snm>Waissbein</snm><fnm>A</fnm></au>
  </aug>
  <source>ACM Transactions on Information and System Security (TISSEC)</source>
  <publisher>ACM</publisher>
  <pubdate>2006</pubdate>
  <volume>9</volume>
  <issue>1</issue>
  <fpage>94</fpage>
  <lpage>-112</lpage>
</bibl>

<bibl id="B17">
  <title><p>Limits on the security of coin flips when half the processors are
  faulty</p></title>
  <aug>
    <au><snm>Cleve</snm><fnm>R</fnm></au>
  </aug>
  <source>Proceedings of the eighteenth annual ACM symposium on Theory of
  computing</source>
  <pubdate>1986</pubdate>
  <fpage>364</fpage>
  <lpage>-369</lpage>
</bibl>

<bibl id="B18">
  <title><p>Zero-knowledge contingent payments revisited: Attacks and payments
  for services</p></title>
  <aug>
    <au><snm>Campanelli</snm><fnm>M</fnm></au>
    <au><snm>Gennaro</snm><fnm>R</fnm></au>
    <au><snm>Goldfeder</snm><fnm>S</fnm></au>
    <au><snm>Nizzardo</snm><fnm>L</fnm></au>
  </aug>
  <source>Proceedings of the 2017 ACM SIGSAC Conference on Computer and
  Communications Security</source>
  <pubdate>2017</pubdate>
  <fpage>229</fpage>
  <lpage>-243</lpage>
</bibl>

<bibl id="B19">
  <title><p>Zerocash: Decentralized anonymous payments from bitcoin</p></title>
  <aug>
    <au><snm>Ben Sasson</snm><fnm>E</fnm></au>
    <au><snm>Chiesa</snm><fnm>A</fnm></au>
    <au><snm>Garman</snm><fnm>C</fnm></au>
    <au><snm>Green</snm><fnm>M</fnm></au>
    <au><snm>Miers</snm><fnm>I</fnm></au>
    <au><snm>Tromer</snm><fnm>E</fnm></au>
    <au><snm>Virza</snm><fnm>M</fnm></au>
  </aug>
  <source>2014 IEEE Symposium on Security and Privacy (SP)</source>
  <pubdate>2014</pubdate>
  <fpage>459</fpage>
  <lpage>-474</lpage>
</bibl>

<bibl id="B20">
  <title><p>Scalable, transparent, and post-quantum secure computational
  integrity</p></title>
  <aug>
    <au><snm>Ben Sasson</snm><fnm>E</fnm></au>
    <au><snm>Bentov</snm><fnm>I</fnm></au>
    <au><snm>Horesh</snm><fnm>Y</fnm></au>
    <au><snm>Riabzev</snm><fnm>M</fnm></au>
  </aug>
  <source>Cryptol. ePrint Arch., Tech. Rep</source>
  <pubdate>2018</pubdate>
  <volume>46</volume>
  <fpage>2018</fpage>
</bibl>

<bibl id="B21">
  <title><p>Succinct non-interactive arguments via linear interactive
  proofs</p></title>
  <aug>
    <au><snm>Bitansky</snm><fnm>N</fnm></au>
    <au><snm>Chiesa</snm><fnm>A</fnm></au>
    <au><snm>Ishai</snm><fnm>Y</fnm></au>
    <au><snm>Paneth</snm><fnm>O</fnm></au>
    <au><snm>Ostrovsky</snm><fnm>R</fnm></au>
  </aug>
  <source>Theory of Cryptography</source>
  <publisher>Berlin, Heidelberg: Springer</publisher>
  <pubdate>2013</pubdate>
  <fpage>315</fpage>
  <lpage>-333</lpage>
</bibl>

</refgrp>
} 


\end{backmatter}

\end{document}